\documentclass[a4paper, amsfonts, amssymb, amsmath, reprint, showkeys, nofootinbib, twoside, preprintnumbers,aps,prd]{revtex4-1}
\usepackage[english]{babel}
\usepackage[utf8]{inputenc}
\usepackage[colorinlistoftodos, color=green!40, prependcaption]{todonotes}
\usepackage{amsthm}
\allowdisplaybreaks
\usepackage{mathtools}
\usepackage{physics}
\usepackage{xcolor}
\usepackage{graphicx}
\usepackage[left=23mm,right=13mm,top=35mm,columnsep=15pt]{geometry} 
\usepackage{adjustbox}
\usepackage{placeins}
\usepackage[T1]{fontenc}
\usepackage{lipsum}
\usepackage{csquotes}
\usepackage{verbatim}
\usepackage[pdftex, pdftitle={Article}, pdfauthor={Author}]{hyperref} 
\usepackage{aas_macros}
\usepackage{multirow}
\begin{document}

\preprint{PI/UAN-2024-727FT}

\title{Neutron stars in the generalized SU(2) Proca theory}
\author{Jhan N. Martínez}
    \email[Correspondence email address: ]{jhamarlo@correo.uis.edu.co}
    \affiliation{Escuela de F\'{\i}sica, Universidad Industrial de Santander,  Ciudad Universitaria, Bucaramanga 680002, Colombia\\}
\author{José F. Rodr\'iguez }
    \email[Correspondence email address: ]{jose.rodriguez2@correo.uis.edu.co}
    \affiliation{Escuela de F\'{\i}sica, Universidad Industrial de Santander,  Ciudad Universitaria, Bucaramanga 680002, Colombia\\}
    \affiliation{ICRANet, Piazza della Repubblica 10, 65122, Pescara PE, Italy\\}
\author{Laura M. Becerra}
    \email[Correspondence email address: ]{laura.becerra@umayor.cl}
    \affiliation{Centro Multidisciplinario de F\'isica, Vicerrector\'ia de Investigaci\'on, Universidad Mayor, Camino La Pir\'amide 5750,  Huechuraba, 8580745, Santiago, Chile\\}
    \affiliation{ICRANet, Piazza della Repubblica 10, 65122, Pescara PE, Italy\\}
\author{Yeinzon Rodr\'iguez }
    \email[Correspondence email address: ]{yeinzon.rodriguez@uan.edu.co}
    \affiliation{Escuela de F\'{\i}sica, Universidad Industrial de Santander,  Ciudad Universitaria, Bucaramanga 680002, Colombia\\}
    \affiliation{ Centro de Investigaciones en Ciencias B\'asicas y Aplicadas, Universidad Antonio Nari\~no, Cra 3 Este \# 47A - 15, Bogot\'a D.C. 110231, Colombia\\}
\author{Gabriel G\'omez}
    \email[Correspondence email address: ]{gabriel.gomez.d@usach.cl}
    \affiliation{Departamento de F\'isica, Universidad de Santiago de Chile,\\Avenida V\'ictor Jara 3493, Estaci\'on Central, 9170124, Santiago, Chile}


\begin{abstract}
The generalized SU(2) Proca theory is a vector-tensor modified gravity theory characterized by an action that remains invariant under both diffeomorphisms and global internal transformations of the SU(2) group. This study aims to further explore the physical properties of the theory within astrophysical contexts. Previous investigations have unveiled intriguing astrophysical solutions, including particle-like configurations and black holes.  The purpose of this work is to constrain the theory's free parameters by modeling realistic neutron stars. To that end, we have assumed solutions that are static, spherically symmetric, and have adopted the t'Hooft-Polyakov magnetic monopole configuration for the vector fields. Employing both analytical techniques, such as asymptotic expansions, and numerical methods involving solving boundary value problems, we have obtained neutron star solutions whose baryonic matter is described by realistic equations of state for nuclear matter. Furthermore, we have constructed mass-radius relations which reveal that neutron stars exhibit greater compactness in comparison with general relativity predictions for most of the solutions we have found and for the employed equations of state. Finally, we have found out solutions where the mass of the star is greater than $\sim$ 2.5 $M_\odot$; this result poses an alternative in the exploration of the mass gap of compact stellar objects.    
\end{abstract}


\maketitle

\section{Introduction} \label{sec:outline}

In 1915, Albert Einstein published the modern vision of the gravitational interaction at the classical level: the theory of General Relativity (GR) \cite{1915SPAW.......844E}. In this theory, gravity is a manifestation of the curvature of spacetime, the latter being inextricably related to the material content of the universe. Tests of this theory, ranging from cosmological to galactic, and, specially, Solar-System scales, have been successful (see e.g. \cite{will2018theory} and references therein).

However, the standard cosmological model, also called the $\Lambda$ Cold Dark Matter model ($\Lambda$CDM), exhibits some shortcomings, particularly regarding the true nature of the exotic fluids known as dark matter and dark energy. It is worth mentioning that one of the first convincing evidences for the existence of dark matter was provided by the flatness of galactic rotation curves, see e.g. \cite{Bertone:2016nfn}. On the other hand, type Ia supernovae have indicated that the universe is experiencing a stage of accelerated expansion today, driven by a repulsive pressure associated commonly with dark energy \cite{amendola2010dark}. Modified classical gravity models, among others, constitute possible solutions to these issues (see e.g. \cite{Capozziello:2011et,Nojiri:2010wj,Nojiri:2017ncd,Heisenberg:2018vsk}). Moreover, besides observational concerns, GR is considered to be just an effective theory \cite{1994PhRvD..50.3874D,Burgess:2003jk} which makes the search for models beyond Einstein's theory a legitimate endeavour.

\textcolor{black}{Among the variety of modified gravity models, we have considered in this paper the generalized SU(2) Proca theory (GSU2P) \cite{GallegoCadavid:2022uzn,GallegoCadavid:2020dho,Allys:2016kbq} (see also \cite{GallegoCadavid:2021ljh} for an extended version). This is a metric theory of gravity that introduces extra vector gravitational degrees of freedom and whose action is invariant under both diffeomorphisms and internal global transformations of the SU(2) group. In addition, the theory is free from the Ostrogradsky instability since it is built so that the field equations are not higher than second order.  The GSU2P is closely related to its Abelian-vector and scalar counterparts, namely the Generalized Proca theory and the Horndeski theory (see e.g. \cite{Rodriguez:2017ckc}). Since a vector inherently induces a preferred direction, vector-tensor theories face challenges in describing the dynamics of a homogeneous and isotropic universe. One possible solution is the cosmic triad configuration (see e.g. \cite{Golovnev:2008cf}). This configuration naturally arises within a theory exhibiting SU(2) symmetry, endowing the GSU2P with a natural framework in the cosmological scenario compared to other vector-tensor theories.}

Despite the phenomenological motivations coming mainly from cosmology \cite{Rodriguez:2017ckc,Rodriguez:2017wkg,Garnica:2021fuu}, the theory must also be tested at astrophysical scales, particularly in the strong field regime where deviations from GR can manifest. A first step was taken in \cite{Martinez:2022wsy}, where soliton solutions were found. In addition, Ref. \cite{2023PhRvD.108b4069G} studied an exact black hole solution. Here, we have proceeded further and derived some consequences for relativistic stars, specifically in the context of neutron stars (NSs). These objects provide an excellent scenario to test deviations from GR. For example, constraints on the mass and radius derived from tidal deformations inferred from the gravitational wave event GW170817 \cite{LIGOScientific:2018cki} and data from the NICER mission \cite{Raaijmakers:2019qny} offer valuable insights. Furthermore, the discovery of \textcolor{black}{very} massive NSs \cite{Demorest:2010bx, Antoniadis:2013pzd, Romani:2022jhd} and exotic objects in the mass gap, $\sim 2.5-5\,M_\odot$ \cite{Ozel:2010su,Fishbach:2020ryj,LIGOScientific:2020zkf,Barr:2024wwl}, raises the natural question of whether these objects emerge  within modified theories of gravity, such as the one studied in this paper.

The purpose of this work is twofold: first, the main one, to put constraints on the free parameters of the theory by constructing static and spherically symmetric neutron star solutions with a realistic equation of state (EOS), and, second, to provide the first hints as to whether the theory can offer possible explanations for the existence of compact objects in the mass gap.

This work is organized as follows. Section \ref{the model} introduces the model and presents both the equations for the structure of a neutron star in the GSU2P theory and the parameterized method for describing the realistic EOS, namely the Generalized Piecewise Polytropic fit \cite{2020PhRvD.102h3027O}. Section \ref{the series} presents analytical solutions of the field equations as asymptotic expansions near the origin and at infinity. In Section \ref{the solutions}, the field equations are solved numerically and mass-radius equilibrium sequences are presented. Finally, Section \ref{the conclusions} discusses the results and suggests directions for future work.

Throughout this article, we use geometrized units ($c = G = 1$), where $c$ is the speed of light and $G$ is the universal gravitational constant. Greek indices represent spacetime indices and run from 0 to 3. Latin indices represent SU(2) group indices and run from 1 to 3. We adopt the sign convention $(+,+,+)$ according to \cite{2017grav.book.....M}.

\section{Generalized SU(2) Proca Theory and Neutron Star solutions} \label{the model}

The action \textcolor{black}{that defines this theory was introduced originally in \cite{Allys:2016kbq} and later reconstructed in \cite{GallegoCadavid:2020dho} (see also \cite{GallegoCadavid:2022uzn}) to take into account the whole constraint algebra that avoids the propagation of unphysical degrees of freedom \cite{ErrastiDiez:2019trb,ErrastiDiez:2019ttn} (see, anyway, \cite{Janaun:2023nxz,ErrastiDiez:2023gme}). In this paper, we have made use of the GSU2P in the form given in \cite{Garnica:2021fuu}:}
\begin{align}
\begin{aligned}
     S = & \frac{1}{16\pi}\int R \ \sqrt{-g}\ d^4x + \frac{1}{16\pi}\int \mathcal{L} \  \sqrt{-g} \ d^4x \\
     & + \int\mathcal{L}_{m}(g_{\mu\nu}, \psi) \ \sqrt{-g} \ d^4x \,,
     \label{eqn:action}
\end{aligned}
\end{align}
where $g$ is the determinant of the metric, $R$ is the Ricci scalar, \textcolor{black}{$\mathcal{L}_{m}$ is the non-gravitational Lagrangian, which depends only on $g_{\mu\nu}$ and the non-gravitational fields represented by $\psi$}, and
\begin{align}
\begin{aligned}
\mathcal{L}=-& F_{\mu \nu}^{a}F^{\mu \nu}_{a}-2\mu^2 B^{\mu}_{a}B_{\mu}^{a} \\
&+\alpha_1\left(\mathcal{L}^1_{4,2}-2\mathcal{L}^4_{4,2}-\frac{20}{3}\mathcal{L}^5_{4,2}+5\mathcal{L}^{7}_{2}\right)\\
&+ \alpha_3\Big(2\mathcal{L}^2_{4,2}+\mathcal{L}^3_{4,2}+\frac{7}{20}\mathcal{L}^4_{4,2}+\frac{14}{3}\mathcal{L}^5_{4,2}\\
&-8\mathcal{L}^6_{4,2}+\mathcal{L}^{7}_{2} \Big)\\
&+ \chi_1 \mathcal{L}^{1}_{2}+\chi_2\mathcal{L}^{2}_{2} +\chi_4\left(\mathcal{L}^{4}_{2}-\frac{\mathcal{L}^{7}_{2}}{2}\right)+\chi_5\mathcal{L}^{5}_{2}\\
&+ \chi_6\left(\mathcal{L}^{6}_{2}-3\mathcal{L}^{7}_{2}\right),
\label{eqn:lagrangian}
\end{aligned}
\end{align}
where $\mu \equiv m_a/\hbar$, $m_a$ being the mass of the vector field $B_\mu^a$ \textcolor{black}{(we have chosen a common value for the three masses) and $\hbar$ being the reduced Planck constant}, $F_{\mu \nu}^{a}\equiv \partial_{\mu}B_{\nu}^{a}-\partial_{\nu}B_{\mu}^{a}+\tilde{g}\epsilon^{a}_{\hspace{2mm}bc}B_{\mu}^{b}B_{\nu}^{c}$ is the gauge field strength tensor, with \textcolor{black}{$\tilde{g}$} being the group coupling constant and \textcolor{black}{$\epsilon^{a}_{\hspace{2mm}bc}$ being the structure
constant tensor of the SU(2) group}, and the $\alpha$'s and $\chi$'s are free parameters of the theory \footnote{In the geometrized units used here, the free parameters of the theory $\alpha_1,\alpha_3, \chi_5, \chi_6$ are dimensionless, while $\chi_1$ and $\chi_2$ have dimensions of $[L]^{-2}$. For convenience, we have redefined all the free parameters by factorizing the global constant $(16\pi)^{-1}$, $\alpha \mapsto (16\pi)^{-1} \alpha, \chi \mapsto (16\pi)^{-1}\chi$. }. The explicit forms of the Lagrangian pieces are given by
\begin{align}
\begin{aligned}
\mathcal{L}^{1}_{4,2}\equiv&\left(B_{b}\cdot B^{b}\right)\left[S^{\mu a}_{\mu}S_{\nu a}^{\nu}-S^{\mu a}_{\nu}S_{\mu a}^{\nu}\right]\\
&+2 \left(B_{a}\cdot B_{b}\right)\left[S^{\mu a}_{\mu}S_{\nu }^{\nu b}-S^{\mu a}_{\nu }S_{\mu }^{\nu b}\right] \,,
\label{}
\end{aligned}
\end{align}
\begin{align}
\begin{aligned}
\mathcal{L}^{2}_{4,2}\equiv&A_{\mu \nu}^{a}S^{\mu b}_{\sigma}B^{\nu}_{a}B^{\sigma}_{b}-A_{\mu \nu}^{a}S^{\mu b}_{\sigma}B^{\nu}_{b}B^{\sigma}_{a}\\
&+A_{\mu \nu}^{a}S^{\sigma b}_{\sigma}B^{\mu}_{a}B^{\nu}_{b} \,,
\label{}
\end{aligned}
\end{align}
\begin{align}
\begin{aligned}
\mathcal{L}^{3}_{4,2}\equiv&B^{\mu a}R^{\alpha}_{\hspace{2mm}\sigma \rho \mu}B_{\alpha a}B^{\rho c}B^{\sigma}_{c}\\
&+\frac{3}{4}\left(B^{a} \cdot B_{a}\right)\left(B_{b} \cdot B^{b}\right)R \,,
\label{eqn:L43}
\end{aligned}
\end{align}
\begin{align}
\begin{aligned}
\mathcal{L}^{4}_{4,2}\equiv&\Big[\left(B^{a}\cdot B_{a}\right)\left(B^{b}\cdot B_{b}\right)\\
&+2\left(B^{a}\cdot B^{b}\right)\left(B_{a}\cdot B_{b}\right)\Big]R \,,
\label{eqn:L44}
\end{aligned}
\end{align}
\begin{equation}
\mathcal{L}^{5}_{4,2} \equiv G_{\mu \nu}B^{\mu a}B^{\nu}_{a}\left(B^{b} \cdot B_{b}\right) \,,
\label{eqn:L45}
\end{equation}
\begin{equation}
\mathcal{L}^{6}_{4,2}\equiv G_{\mu \nu}B^{\mu a}B^{\nu b}\left(B_{a} \cdot B_{b}\right) \,,
\label{eqn:L46}
\end{equation}
\begin{equation}
\mathcal{L}^{1}_{2} \equiv (B^{a} \cdot B_{a}) (B^{b} \cdot B_{b}) \,,
\label{}
\end{equation}
\begin{equation}
\mathcal{L}^{2}_{2} \equiv (B^{a} \cdot B_{b}) (B^{b} \cdot B_{a}) \,,
\label{}
\end{equation}
\begin{equation}
\mathcal{L}^{3}_{2} \equiv B_{\mu}^{b} B_{\rho b} A^{\mu \nu a} A^{\rho}_{\;\; \nu a} \,,
\label{}
\end{equation}
\begin{equation}
\mathcal{L}^{4}_{2} \equiv B_{\mu}^{b} B_{\rho a} A^{\mu \nu a} A^{\rho}_{\;\;\nu b} \,,
\label{}
\end{equation}
\begin{equation}
\mathcal{L}^{5}_{2} \equiv B_{\mu a} B_{\rho}^{b} A^{\mu \nu a} A^{\rho}_{\;\; \nu b} \,,
\label{}
\end{equation}
\begin{equation}
\mathcal{L}^{6}_{2} \equiv (B^{b} \cdot B_{b}) A_{\mu \nu a} A^{\mu \nu a} \,,
\label{}
\end{equation}
\begin{equation}
\mathcal{L}^{7}_{2} \equiv (B^{b} \cdot B_{a}) A_{\mu \nu b} A^{\mu \nu a} \,,
\label{}
\end{equation}
where $A^{a}_{\mu \nu} \equiv \nabla_{\mu}B_{\nu}^{a} - \nabla_{\nu}B_{\mu}^{a}$ and $S^{a}_{\mu \nu} \equiv \nabla_{\mu}B_{\nu}^{a} + \nabla_{\nu}B_{\mu}^{a}$ \textcolor{black}{are, respectively, the Abelian version of the non-Abelian gauge 
field strength tensor and the symmetric version of $A^{a}_{\mu \nu}$,} \textcolor{black}{$R^\alpha_{\hspace{2mm} \sigma \rho \mu}$ is the Riemann tensor, and $G_{\mu \nu}$ is the Einstein tensor.} \textcolor{black}{The theory is clearly formulated in the Jordan frame since the vector field is non-minimally coupled to gravity in the Lagrangian pieces $\mathcal{L}_{4,2}^3, \mathcal{L}_{4,2}^4, \mathcal{L}_{4,2}^5$, and $\mathcal{L}_{4,2}^6$;  $B_\mu^a$ is considered as the only one extra gravitational degree of freedom, when comparing this theory to GR, and, therefore, it is reasonable to minimally couple regular matter to gravity.}

The explicit form given by Eq. \eqref{eqn:lagrangian} for the Lagrangian was derived to ensure that tensor perturbations on a Friedmann-Lemaître-Robertson-Walker background match those in GR at least up to second order \cite{Garnica:2021fuu}.\footnote{This is, of course, one possibility among others within the more general condition that the action may behave in a different way compared to GR, at the perturbative level, but still avoiding instability problems \cite{Gomez:2019tbj}.} This results in luminal gravitational wave propagation and freedom from ghosts and Laplacian instabilities in the tensor sector. Compliance with the conditions imposed by the gravitational wave event GW170817 \cite{LIGOScientific:2017vwq} and its optical counterpart GRB 170817A \cite{LIGOScientific:2017zic} supports the  plausibility of the theory at cosmological scales. However, the extension of this constraint to astrophysical scales is uncertain as it depends on the specific background configuration. Determining whether luminal propagation persists near astrophysical objects requires further study, including obtaining and perturbing the background solution, which will be addressed in a future work. However, for the sake of simplicity, we have assumed that the validity of those results extends to the scale of compact objects.

\subsection{Field equations}
The field equations have been obtained by varying the action \eqref{eqn:action} with respect to $g^{\mu\nu}$ and $B^{a\mu}$, respectively:
\begin{equation}
G_{\mu \nu}=8\pi (T_{\mu \nu}^{\rm v} +T_{\mu\nu}^{m})\,,
\label{eqn:GNAP}
\end{equation}
\begin{equation}
K_{a \mu} \equiv\frac{1}{4}\frac{\delta \mathcal{L}}{\delta B^{a\mu}} =  0 \,,\label{eqn:Kamu}
\end{equation}
where
\begin{equation}
T_{\mu \nu}^{\rm v}\equiv -\frac{1}{8\pi \sqrt{-g}}\frac{\delta (\mathcal{L}\sqrt{-g})}{\delta g^{\mu \nu}} \,.
\label{eqn:Tv}
\end{equation}
\begin{equation}
T_{\mu \nu}^{m}\equiv -\frac{1}{8\pi \sqrt{-g}}\frac{\delta (\mathcal{L}_{m}\sqrt{-g})}{\delta g^{\mu \nu}} \,
\label{eqn:Tm}
\end{equation}
Appendix A in Ref. \cite{Martinez:2022wsy} contains the explicit forms of (\ref{eqn:Kamu}) and (\ref{eqn:Tv}).  

The vector fields represent additional degrees of freedom of the gravitational interaction and can be interpreted as a dark fluid. This interpretation facilitates the definition of an effective energy-momentum tensor, $ T^{\rm v}_{\mu\nu}$, associated with the SU(2) vector field. However, it is worthwhile emphasising that \emph{the vector fields are fundamentally gravitational degrees of freedom}. The dark fluid interpretation is merely used to compare the contributions from the vector fields with those from baryonic matter. Accordingly, we have defined the energy density and isotropic pressure associated with the vector fields, measured by an observer whose four-velocity is $u^{\mu}$, as
\begin{equation}
\epsilon_{\rm v}\equiv T_{\mu \nu}^{\rm v}u^{\mu}u^{\nu} \,,
\label{eqn:rho}
\end{equation}
\begin{equation}
P_{\rm v}\equiv\frac{1}{3}(g^{\mu \nu}+u^{\mu}u^{\nu})T_{\mu \nu}^{\rm v} \,,
\label{eqn:p}
\end{equation}
respectively.
\subsection{Static and spherically symmetric equations}
We now focus on the static and spherically symmetric case in order to describe isolated and non-rotating NSs.
The line element of a static and spherically symmetric spacetime in Schwarzschild coordinates is given by
\begin{equation}
ds^2 = -e^{2\Phi(r)}dt^2 + h(r)^{-1} dr^2 + r^2(d\theta^2 +\sin^2\theta d\phi^2) \,, \label{eqn:metricII}
\end{equation}
where $h(r)\equiv 1-2m(r)/r$. Given the structure of the field equations, we have found it more convenient to use the variable $e^{-2\delta}\equiv e^{2\Phi}/h$.

The configuration we have chosen for the vector field $B_{\mu}^a$ is given by the t'Hooft-Polyakov monopole \emph{ansatz} which is a special case of the more general spherically symmetric one \cite{1977PhRvL..38..121W} (see also Refs. \cite{Sivers:1986kq,Forgacs:1979zs}):
\begin{equation}
    \mathbf{B}= -\frac{\tau^a}{\tilde{g}}\biggl[ \epsilon_{ajk}x^k\frac{(1+w)}{r^2}dx^j\biggr] \,,
\end{equation}
where $\tau^a = -i\sigma^a/2$ is a basis for the SU(2) algebra with $\sigma^a$ being the Pauli matrices, $x^a$ (or $x^j$) are the space-time cartesian coordinates, and $\epsilon_{aij}$ is the Levi-Civita tensor.
The cartesian coordinates $x^i$ can be obtained from the polar spherical coordinates $(r, \theta,\phi)$ in the same way as in flat space \cite{1988PhRvL..61..141B, 1993PhRvD..47.2242G}. Thus, the t'Hooft-Polyakov monopole has the following more convenient form:
\begin{equation}
    \mathbf{B}= (1+w)(\tau_\theta \sin\theta d\phi -\tau_\phi d\theta) \,, \label{eqn:poly}
\end{equation}
where  
\begin{align}
\tau_r &= \sin\theta \cos\phi\, \tau_1 + \sin\theta\sin\phi \, \tau_2 + \cos\theta \, \tau_3\,,\\
\tau_\theta &= \cos\theta\cos\phi \, \tau_1 + \cos\theta\sin\phi\, \tau_2 - \sin\theta \, \tau_3\,,\\
\tau_\phi  &= -\sin\phi \, \tau_1 + \cos\phi\, \tau_2 \,,
\end{align}
is a polar spherical basis for the SU(2) algebra.

\textcolor{black}{In this study, we have assumed such a specific configuration for the vector field because it technically simplifies our analysis. In more general cases involving additional functions of the vector field beyond $w$, the resultant field equations are generally not independent. This will require further analysis to eliminate redundancy in the system. Such complexities can arise, indeed, even in standard Yang-Mills theory. In contrast, the pure magnetic ansatz presented in Eq. (\ref{eqn:poly}) naturally leads to a very simple set of field equations\footnote{\textcolor{black}{We individually tested each function of the most general spherically symmetric configuration, and the t’Hooft-Polyakov configuration was the only one that led to a physically consistent system of equations.}}. 
From an astrophysical perspective, this configuration has shown promise in contexts such as solitons \cite{Martinez:2022wsy} and black holes \cite{2023PhRvD.108b4069G}, yet it remains underexplored in the context of neutron stars.}

\textcolor{black}{All non-gravitational fields couple only to the metric tensor, in a minimal way, in accordance with the Einstein Equivalence Principle (minimal coupling prescription).} For the NS, we have considered a single perfect fluid, $(T^{m})^{\mu}{}_{\nu}={\rm diag}(-\epsilon(r), P(r), P(r), P(r))$, 
where $\epsilon$ and $P$ are the energy density and pressure, respectively, measured by a static observer with respect to the fluid. The rest mass density of baryonic matter is denoted by $\rho$.
The equation of motion for matter is given by the continuity equation, $\nabla_{\nu} T_m^{\mu\nu}=0$, which reduces to
\begin{equation}
     P' = - (P + 
\epsilon)\Phi' = (P + 
\epsilon)\biggl(\frac{m - rm'}{2mr-r^2}+\delta'\biggr) \,,
\end{equation}
\textcolor{black}{where the prime means a derivative with respect to $r$.}

The  equations in the static and spherically symmetric case, together with the t'Hoof-Polyakov \emph{ansatz}, have a rather complicated form which can be found in Appendix \ref{appES}. 

In order to obtain objects that \textcolor{black}{correspond to} isolated solutions, the asymptotically flatness condition must be satisfied. Thus, we \textcolor{black}{have} imposed boundary conditions at spatial infinity $r\to\infty$ \textcolor{black}{so that} $\Phi\to 0, \delta\to 0$, and $m\to{\rm const.}$ \textcolor{black}{The radius of the NS is obtained by requesting the condition
\begin{equation}
    P(R)= 0 \,,
\end{equation}
and by requiring that both the pressure and the energy density of the baryonic matter vanish when $r>R$. This does not imply by any means, however, that the vector fields outside the NS get a trivial value.} 

\textcolor{black}{We have chosen the Komar mass \cite{1963PhRv..129.1873K} to define the gravitational mass. This mass, which represents the conserved charge associated to the time-like Killing vector \cite{wald2010general}, is given by the asymptotic value of the mass function (see Appendix \ref{sec:appKomarADM})}
\begin{equation}
    M = \lim_{r\to\infty} m(r) = \lim_{r\to \infty}\frac{r}{2}(1-h) \,. \label{km}
\end{equation}
\textcolor{black}{When the vector field $B_\mu^a$ is minimally coupled to the metric $g_{\mu\nu}$, the Komar mass and the Arnowitt–Deser–Misner (ADM) mass \cite{gourgoulhon20123+1}, the latter being the on-shell value of the Hamiltonian, might coincide. On the other hand, these two masses do not coincide in the non-minimal coupling case due to the presence of additional boundary terms (see Appendix  \ref{sec:GYH} for details).}

\textcolor{black}{In addition, following \cite{1988PhRvL..61..141B}, we have defined an effective charge,
\begin{equation}
Q_{\rm eff}^2 \equiv 2r(M-m) \,.
\label{eqn:Qeff}
\end{equation}
This charge does not represent the strength of an interaction, it is only used to quantify the departure of the metric solution from the Schwarzschild one.}

Finally, the group coupling constant $\tilde{g}$ has inverse units of length \textcolor{black}{in geometrized units}. Accordingly, we \textcolor{black}{have} defined the normalized variables $\hat{r} \equiv r\tilde{g}$, $\hat{m}\equiv
 m \tilde{g}$, $\hat{P}\equiv P\tilde{g}^{-2}$ and $\hat{\epsilon} \equiv \epsilon\tilde{g}^{-2}$.  Since we are interested in astrophysical scales, we have set the unit length to $\tilde{g}^{-1} \equiv M_\odot = 1.477$~km. 

\subsection{Equations of state \textcolor{black}{for the} NS}\label{sec:EOS}

For this paper, we have explored baryonic matter  EOSs that  are able to model objects with more than $2~M_\odot$ in GR and agree with the data from NICER \cite{Riley:2019yda,Riley:2021pdl}.  Thus, in this context, we have allowed for the possibility of having more massive configurations within this theory.  Among the NS EOSs in the literature, we have chosen to work with the H4 \cite{Lackey:2005tk} and the GM1Y6 \cite{1991PhRvL..67.2414G,Oertel:2014qza} EOSs. Both  EOSs correspond to matter formed by electrons, nucleons,  and hyperons.  

For simplicity in our numerical calculations,  these EOSs  are parameterized using the Generalized Piecewise Polytropic  fit \cite{2020PhRvD.102h3027O}. We divide  the baryonic rest mass density range in $N$ intervals.  Within each interval, spanning from $\rho_i$ to $\rho_{i+1}$, the pressure and energy density are:
\begin{eqnarray}\label{eq:poly_eos}
    P(\rho) &=& K_i\rho^{\Gamma_i} + \Lambda_i \,,\\
    \epsilon(\rho) &=& \frac{K_i}{\Gamma_i-1}\rho^{\Gamma_i} +(1+a_i)\rho - \Lambda_i \,.
\end{eqnarray}
The polytropic indices, $\Gamma_i$, and the dividing densities, $\rho_i$, are the fit parameters of the method.  For the high-density region of the EOSs (densities greater than the nuclear saturation density, $\rho_s\approx 2.4 \times 10^{14}$ g cm$^{-3}$), we have used a three-zone Piecewise Polytropic model  (see Table~\ref{tab:EoS}; for more details, see \cite{2024PhRvD.109d3025B}), while for the low-density region ($\rho<\rho_s$), we have employed the five zone parametrization presented in \cite{2020PhRvD.102h3027O}.

 The  values of the constants $K_i$, $\Lambda_i$ and $a_i$ are established by ensuring the continuity  of the energy density, pressure and sound speed at the  dividing densities:
\begin{eqnarray}
    K_{i+1}&=& K_i\frac{\Gamma_i}{\Gamma_{i+1}}\rho^{\Gamma_i-\Gamma_{i+1}} \,,\label{eq:Ki}\\
    \Lambda_{i+1}&=& \Lambda_i + \left(1-\frac{\Gamma_i}{\Gamma_{i+1}}\right)K_i\rho^{\Gamma_i} \,,\label{eq:Lambdai}\\
    a_{i+1}&=& a_i +\Gamma_i\frac{\Gamma_{i+1}-\Gamma_i}{(\Gamma_{i+1}-1)(\Gamma_i-1)}K_i\rho^{\Gamma_i-1} \,.\label{eq:ai}
\end{eqnarray}
 
The above recurrence relations are solved by setting the parameters in the lowest density zone to   $a_8=0$, $\Lambda_8=0$ and $K_8=5.21 \times 10^{-9} c^2$ in cgs units, where $c$ is the speed of light. \textcolor{black}{The reported RMS error between the fitted EOS and the actual EOS is less than 1\% \cite{2020PhRvD.102h3027O,2024PhRvD.109d3025B}.  Both the GM1Y6 and H4 EOSs can represent physical objects with subliminal sound speed (see \cite{2024PhRvD.109d3025B} for more details).}

\begin{table}
    \begin{tabular}{c|cccccc}\hline
        $\quad$ \textbf{EOS}  &  $\mathbf{\log \rho_1} $  & $ \mathbf{\log \rho_2} $  &  $ \mathbf{\log \rho_3} $  & $\Gamma_1$ & $\Gamma_2$ & $\Gamma_3$   \\
          &$[\mathrm{g\, cm^{-3}}]$  &$[\mathrm{g\, cm^{-3}}]$ &  $[\mathrm{g\, cm^{-3}}]$ &  &  &   \\
         &  &  &  &  &  &   \\\hline
        
        H4 & $14.99$ &  $14.87$ & $13.49$ & $2.51$	& $2.33$ &	$1.56$   \\
        GM1Y6  & $14.99$ &  $14.87$ & $13.75$ & $2.96$ &$1.62$ &$1.63$ \\
       
        &  &  &  &  &  &  \\
        \hline
    \end{tabular}
    \caption{ High-density  Generalized Piecewise Polytropic  fit parameters for the H4 and GM1Y6 EOSs.}   \label{tab:EoS}
\end{table}
%

\section{Asymptotic series solution at  the origin and at spatial infinity} \label{the series}
\begin{figure*}[!t]
    \centering
    \includegraphics[width=0.48\textwidth]{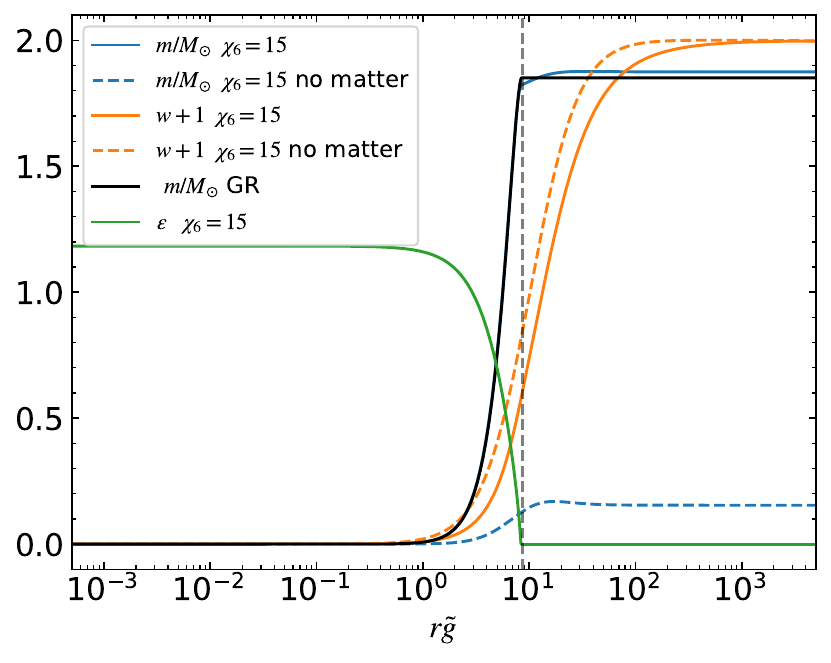}
    \includegraphics[width=0.48\textwidth]{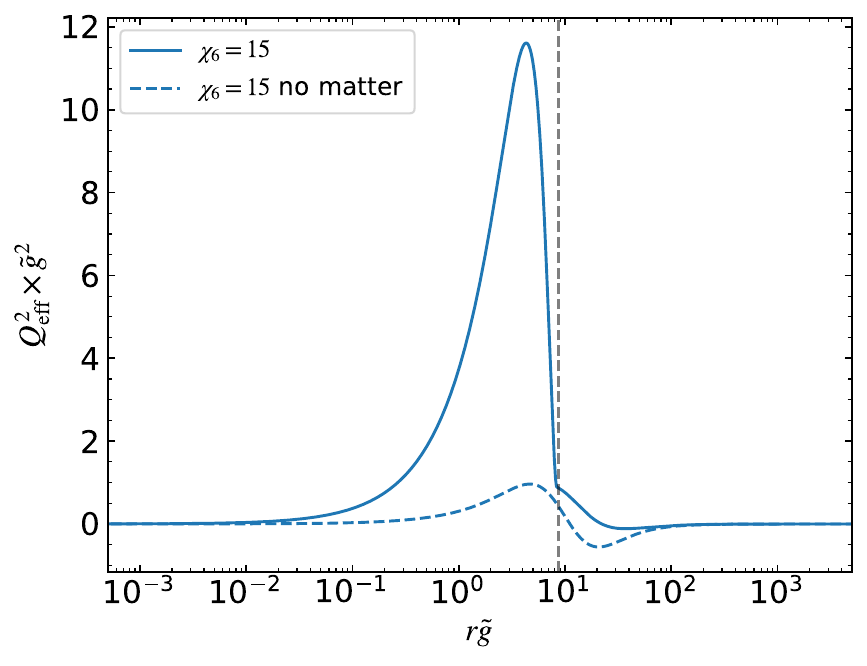}
    \caption{Left: example of a 1-node solution in the minimal coupling case $\chi_6=15$. We have used here the EOS H4 and central rest-mass density $\rho_0 = 1\times10^{15}$~g~cm$^{-3}$.  The mass function $m$ is shown in blue (solid line) and the vector field $w$ in orange (solid line). The NS \textcolor{black}{mass function} in GR is shown as a solid black curve. The GSU2P solution departs from the GR mass function at a radius \textcolor{black}{$r\tilde{g} \sim \mathcal{O}(1) $} where the vector field starts to increase. For comparison and contrast, we  also show the particle-like (no baryonic matter) solution, with dashed lines, employing the chosen value for $\chi_6$. For the case $\chi_6=15$, the energy density of baryonic matter  is shown in green and the radius of the NS as the vertical dashed grey line. Right: effective charge \eqref{eqn:Qeff} for the GSU2P solutions shown in the left panel. \textcolor{black}{In both panels, the grey dashed vertical line marks the radius of the NS.}}
    \label{fig:chi6-30-0047}
\end{figure*}
In order to find a regular solution at the origin, we have expanded the field variables as asymptotic power series:
\begin{align}
\hat{m} &=a_3\hat{r}^3 + a_5\hat{r}^5   +\mathcal{O}(\hat{r}^6) \,, \label{eqn:mseries0}\\
w &= -1 +b_2 \hat{r}^2 + b_4 \hat{r}^4 +\mathcal{O}(\hat{r}^6) \,, \label{eqn:wseries0}\\
\Phi &= 1 + c_2 \hat{r}^2 + c_4 \hat{r}^4 +\mathcal{O}(\hat{r}^6) \,, \label{eqn:Phiseries}\\
\hat{P} &= \hat{P}_0+d_2\hat{r}^2+d_4\hat{r}^4+\mathcal{O}(\hat{r}^6)\,,
\label{eqn:Pseries}
\end{align}
where $\hat{P}_0$ is the normalized central pressure of the baryonic matter. The different coefficients have been found by solving the equations of structure \eqref{eqn:eqm-min}-\eqref{eqn:eqw-nonmin}. The explicit value of the non-vanishing coefficients can be found in the Appendix \ref{sec:appSeries}. The series depend on both $b_2$ and on the value of the pressure of the baryonic matter at the centre. 

The coefficient $b_2$ is arbitrary and is related to the central normalized energy density associated to the vector fields, $2 \pi\hat{\epsilon}_{\rm v}(0) = 3b_2^2$.  The coefficient $d_1$ vanishes and $d_2<0$, thus the pressure reaches a (local) maximum at the origin, as expected. The leading contributions in the series come from the baryonic matter and from the Einstein-Yang-Mills (EYM) theory. Modifications in the behaviour of the field variables are expected to appear in the intermediate region $\hat{r}\sim 1$ where the vector fields are not trivial. The pressure and density of the baryonic matter is also seen to be modified explicitly at higher orders by the vector fields which shows that the new gravitational degrees of freedom indeed alter the baryonic matter distribution in the star.

There is no an equivalent to the Birkhoff's theorem in the GSU2P theory. Consequently, we must search for the possible vacuum asymptotically flat solutions. In order to achieve this task, we have expanded the variables at spatial infinity as asymptotic series in inverse powers of $r$: 
\begin{align}
\hat{m} &= \hat{M}+\frac{\tilde{a}_1}{\tilde{r}}+\frac{\tilde{a}_2}{\hat{r}^2}+\frac{\tilde{a}_3}{\hat{r}^3}+\mathcal{O}\left(\frac{1}{\hat{r}^4}\right) \,,
\label{eqn:m-asymp-infty}\\
w &= w_\infty+\frac{\tilde{b}_1}{\hat{r}}+\frac{\tilde{b}_2}{\hat{r}^2}+\frac{\tilde{b}_3}{\hat{r}^3}+\mathcal{O}\left(\frac{1}{\hat{r}^4}\right) \,,
\label{eqn:w-asymp-infty}\\
\Phi &= \Phi_\infty+\frac{\tilde{c}_1}{\hat{r}}+\frac{\tilde{c}_2}{\hat{r}^2}+\frac{\tilde{c}_3}{\hat{r}^3}+\mathcal{O}\left(\frac{1}{\hat{r}^4}\right) \,,
\end{align}
and have solved the field equations. The solutions have the same \emph{form} as in the particle-like case (see  \cite{Martinez:2022wsy} for details) with the same free parameters (which are determined later by numerical integration). Therefore, there exist asymptotically flat solutions as required. It remains to be seen if we can answer the question of whether these solutions can be matched with the found out solutions near the origin. For this task, we have proceeded to numerically solve the equations of structure by a shooting method with parameter $b_2$. The two asymptotic solution are matched in an intermediate region numerically and the free parameters of both asymptotic series are determined. 

Particle-like solutions and NS solutions share the same form of the asymptotic series at infinity but the free coefficients are \textcolor{black}{different}, e.g. the gravitational mass is different. This comes from the fact that the behaviour of the field variables is different in the intermediate region due to the presence of baryonic matter (see Fig. \ref{fig:chi6-30-0047}). 
\section{1-node  neutron star solutions} \label{the solutions}

We have solved the structure equations using a shooting method with the parameter $b_2$ to match the asymptotic solution at infinity. Given the complexity of the GSU2P theory, we have selected two representative cases in the Lagrangian. One case involves the vector fields being minimally coupled to the metric where $\chi_6\neq0$ and all other parameters vanish. The other case involves a non-minimal coupling where $\alpha_1\neq0$ and all other parameters vanish. The different solutions can be classified by the number of nodes in the field variable $w$. We have focused on the 1-node solutions since they are potentially stable in the particle-like case \cite{Martinez:2022wsy}.

Before starting the numerical integration of the structure equations in the GSU2P theory, we conducted exploratory work in the standard EYM case which, to the best of our knowledge, had not been studied so far. Interestingly, we have found out solutions in the presence of baryonic matter. However, these solutions turned out to be unrealistic, consisting of a very small core of baryonic matter surrounded by an extended halo made of YM fields. These objects resemble the EYM soliton with a very compact baryonic mass core.  For these objects, the energy density of the (gravitational) vector fields is dominant over the non-gravitational fields. This phenomenon was also observed at cosmological scales \cite{Santiago}. Therefore, to construct realistic NS models, we searched for cases within the GSU2P theory where the YM energy was screened by the other Lagrangian terms. \textcolor{black}{Based on the above discussion, we have not reported solutions in the EYM case here because they are not of astrophysical interest.}

\subsection{Minimal coupling case}
\begin{figure*}[!t]
    \centering
    \includegraphics[width=0.49\textwidth]{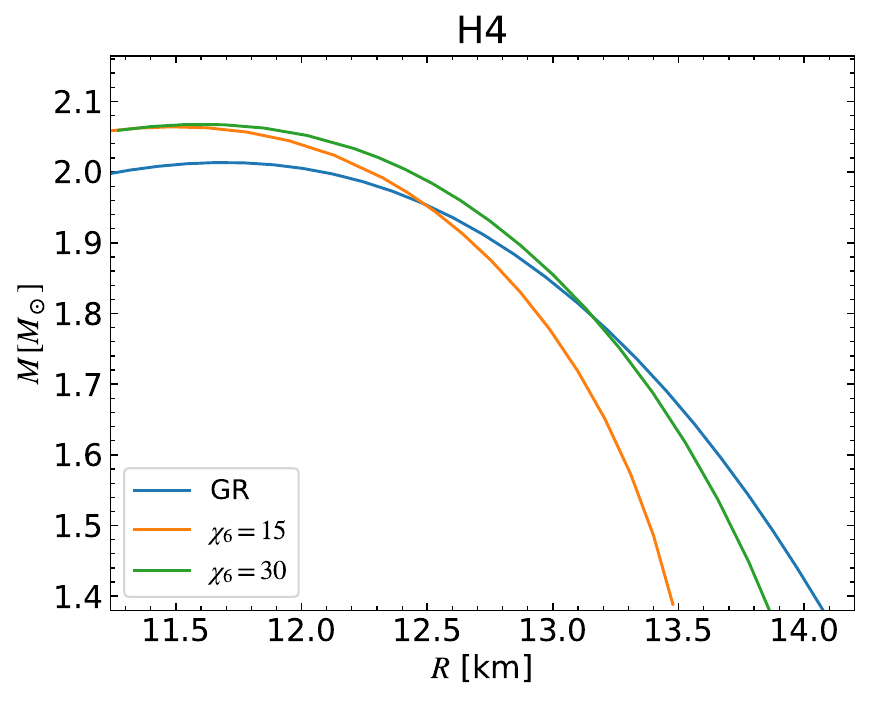}
    \includegraphics[width=0.49\textwidth]{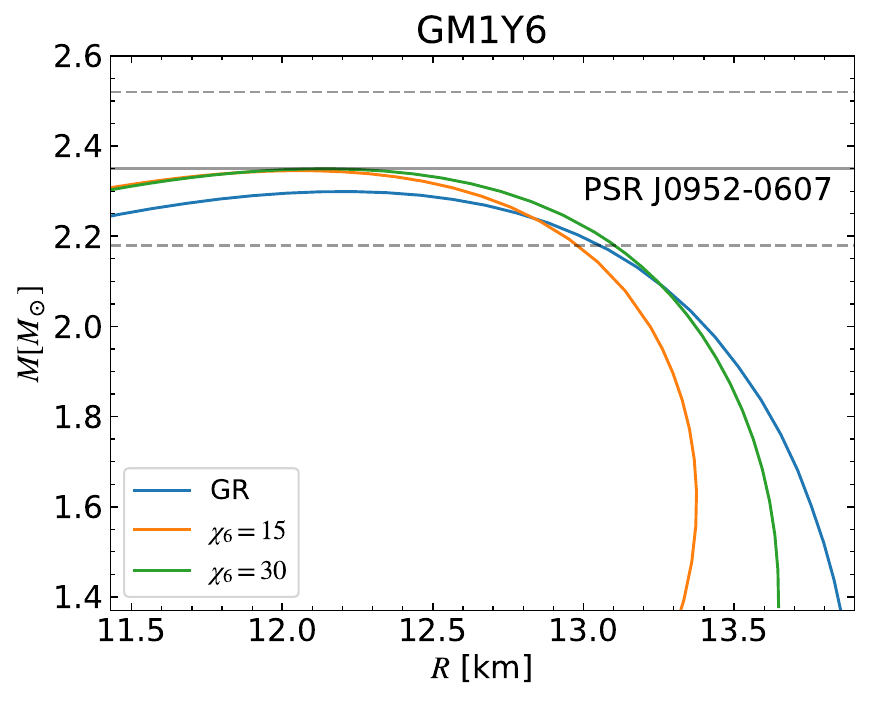}  
    \caption{Gravitational mass versus radius of a NS whose baryonic matter is described by the EOS H4 (left) and GM1Y6 (right) for the minimal case $\chi_6$. The horizontal grey solid line in the right panel corresponds to the mass of the potentially heaviest observed NS PSR J0952-0607 ($2.35\pm 0.17 M_\odot$) \cite{Romani:2022jhd}, the $1
\sigma$ uncertainty being represented by dashed grey lines.}
    \label{fig:MRChiC6}
\end{figure*}

We have aimed at identifying cases where the YM energy is at least comparable to the central energy density of baryonic matter. To achieve this, as a first approximation, we have used an EOS described by a single polytropic equation. For the minimal coupling case, we have found that $\chi_6 \gtrsim 10$ fulfils this requirement.

\textcolor{black}{Next, we have employed a realistic EOS with the parameterization described in Subsection \ref{sec:EOS}. In Fig. \ref{fig:chi6-30-0047}, we show the behaviour of the mass function $m$ and the vector field $w$ for the specific case $\chi_6=15$ with EOS H4 and central rest-mass density $\rho_0 = 1\times10^{15}$~g~cm$^{-3}$. The mass function is seen to deviate from the GR case from the intermediate region \textcolor{black}{$r\tilde{g}\sim \mathcal{O}(1)$} where the vector field value is non-trivial, $w\neq \pm1$. The mass function continues to increase even though the energy density of baryonic matter has vanished, see the green curve in the left panel of Fig. \ref{fig:chi6-30-0047}. This occurs because the energy density of the vector fields does not vanish at the same radius as the baryonic matter does; instead, it decreases more slowly outside the star, keeping its contribution to the gravitational mass.}

\textcolor{black}{In Fig. \ref{fig:chi6-30-0047}, we have also compared the NS solution and the one with no baryonic matter, i.e., the particle-like solution using the same value for $\chi_6$ for comparison. The mass of the particle-like solution is seen to be much less than that of the NS in both the GSU2P theory and GR. Nevertheless, the mass of the NS in the GSU2P is greater than in GR because of the contribution of the vector fields, although this is not a linear effect as can be appreciated in the left panel of the same figure.} 
In addition, the presence of baryons increases notably the value of the effective charge\textcolor{black}{\footnote{\textcolor{black}{A curious reader might ask why the square of the effective charge becomes negative in the absence of matter, specifically around $r\tilde{g}\sim 10^{2}$ . This is due to the dominance of repulsive interactions between the vector fields. Additionally, this charge has a topological origin, unlike the electromagnetic charge which is associated with a Noether current.}}}, see equation \eqref{eqn:Qeff} and the right panel of Fig.~\ref{fig:chi6-30-0047}. \textcolor{black}{There is a noticeable kink around $r \tilde{g} \approx 10$, which corresponds to the point where $P=0$, i.e., the radius of the NS, $R$. This kink is produced at the edge of the NS. Beyond this point, the energy density associated with the vector field continues to contribute to the charge, although its contribution becomes much smaller compared to that of the baryonic matter. For values of $r \tilde{g}$ larger than $R$, the effective charges in both cases do not coincide due to the nonlinear nature of the theory. Furthermore, the square of the effective charge vanishes in the asymptotic region, indicating that these configurations are globally neutral.}

\begin{figure*}
    \centering
    \includegraphics[width=0.9\linewidth]{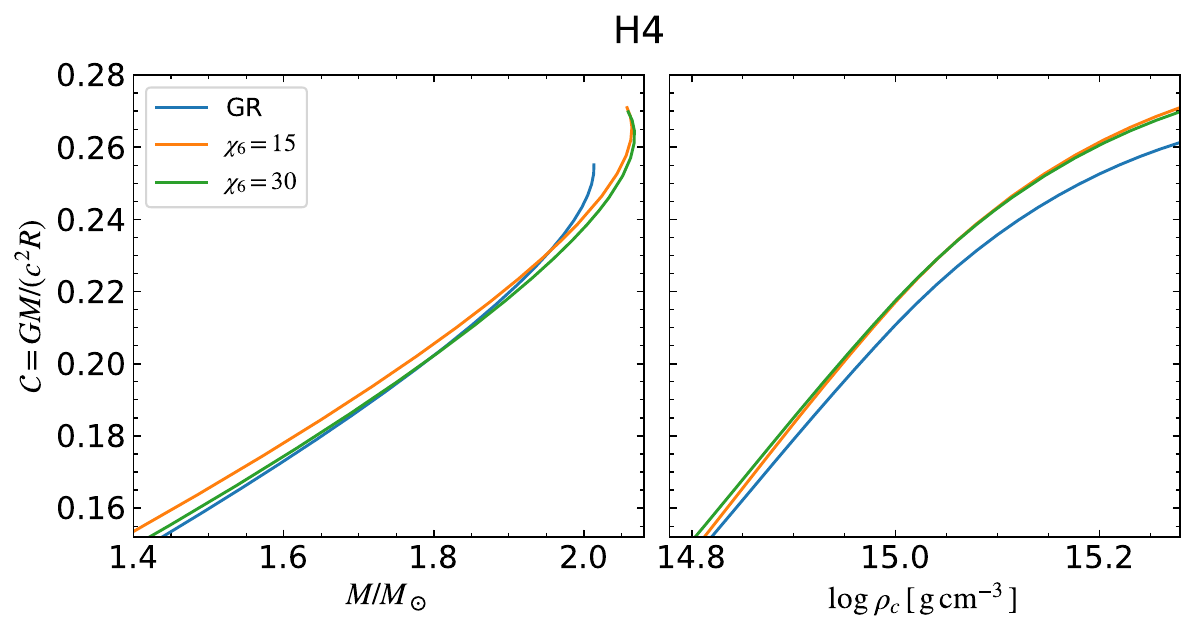}
    \caption{\textcolor{black}{Compactness of the H4 EOS equilibrium configurations shown in the left panel of Fig. \ref{fig:MRChiC6} as a function of the gravitational mass (left) and of the central baryonic mass density (right).}}
    \label{fig:CompactnessC6}
\end{figure*}

In Fig. \ref{fig:MRChiC6}, we show the equilibrium sequence of mass versus radius for two different values of the parameter $\chi_6$, and for the EOS H4 (left panel) and EOS GM1Y6 (right panel). The maximum mass is greater than in the GR case. When the central density of the baryonic mass decreases, the contribution of the vector fields increases and the radius becomes smaller. As a consequence, as the gravitational mass decreases, the radius becomes smaller, and there is a tendency to create a very compact core of baryonic matter as in the EYM case. \textcolor{black}{For the H4 EOS, we show in Fig. \ref{fig:CompactnessC6} the compactness, $\mathcal{C} = M/R$, of the equilibrium configurations as a function of the gravitational mass and of the central baryonic mass density. When we compare configurations with the same gravitational mass, for low-mass cases, the GSU2P solution is more compact than the GR solution, as seen in the left panel of Fig. \ref{fig:CompactnessC6}. This reafirms that for low central baryonic mass density, the contribution of vector fields becomes important and creates a more compact object, as observed in the right panel of Fig. \ref{fig:CompactnessC6}. Additionally, when we compare configurations with the same central baryonic mass density, the GSU2P solution is always more compact. The baryonic matter predominantly contributes to the mass whereas the vector fields make a more subtle contribution. However, this subtle contribution makes the GSU2P solutions more compact.}

In the GR case, \textcolor{black}{the selected EOSs} exhibit a maximum mass that is less than the central value of the mass of the observed pulsar PSR J0952-0607, but within $1\sigma$ uncertainty. In the GSU2P case, the maximum mass is closer to this central value of the heaviest observed NS.
\begin{figure}
    \centering
\includegraphics[width=0.48\textwidth]{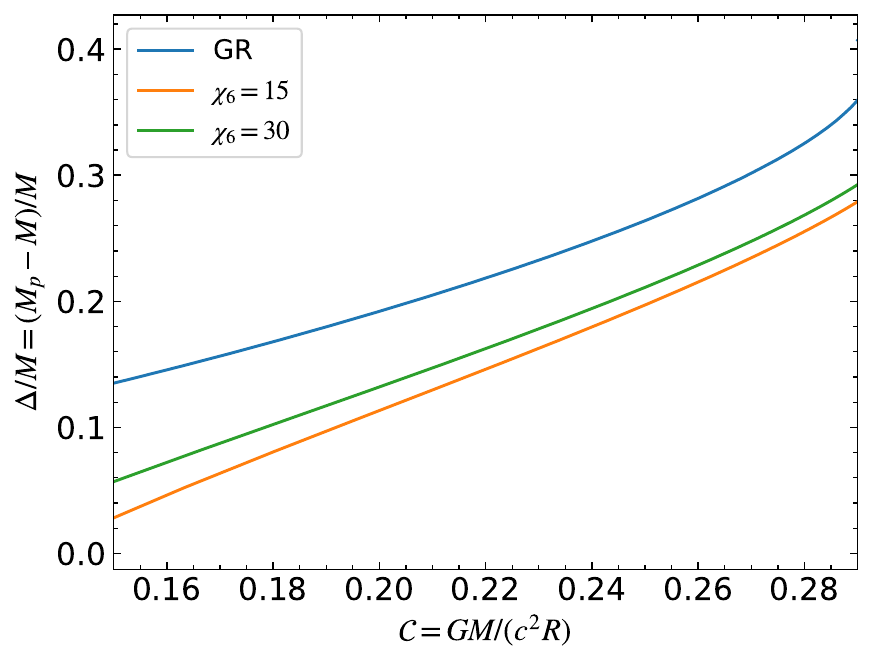}
    \caption{Specific gravitational binding energy $\Delta/M = (M_p - M)/M$ for the equilibrium sequence shown in the right panel of Fig. \ref{fig:MRChiC6}.}
    \label{fig:Delta}
\end{figure}

In order to assess the stability of the star, we have computed the proper mass-energy
 \begin{equation}
     M_p \equiv \int_0^{r_s} \frac{4\pi r^2 \epsilon}{\sqrt{1-2 m(r)/r}}dr \,, 
 \end{equation}
and the gravitational binding energy, $\Delta \equiv M_p - M$ \cite{1959ApJ...130..884C}. We have found that this last value is generally positive, see Fig. \ref{fig:Delta}, satisfying the minimum necessary condition for stability. Nevertheless, \textcolor{black}{to properly check the stability, the explicit perturbation theory must be carried out which will be left for a future work.} When the effective energy density associated to the vector fields (tends to) become(s) dominant, $\Delta$ tends to become negative, i.e., when the energy density is dominated by the vector fields, especially by the YM term, the configuration can be unstable as expected from the particle-like EYM solutions.

\textcolor{black}{Finally, for this minimally coupled case, we have explored evidence suggesting the possible existence of objects within the mass gap of $2.5\, M_\odot <M<5\,M_\odot$. We have specifically searched for NS solutions with masses greater than those we have found out previously, while maintaining a similar level of compactness. To achieve this and \emph{only here,} we have adjusted the value of the gauge coupling constant. As previously mentioned, the gauge coupling constant represents the mass-length scale of the system. Altering the value of $\tilde{g}$ changes the mass and length of the configurations. However, this change \emph{is not merely a rescaling factor} due to the presence of baryonic matter. For particle-like solutions, modifying the group constant does result in a rescaling of the solutions.}  

\begin{figure}
    \centering
    \includegraphics[width=0.49\textwidth]{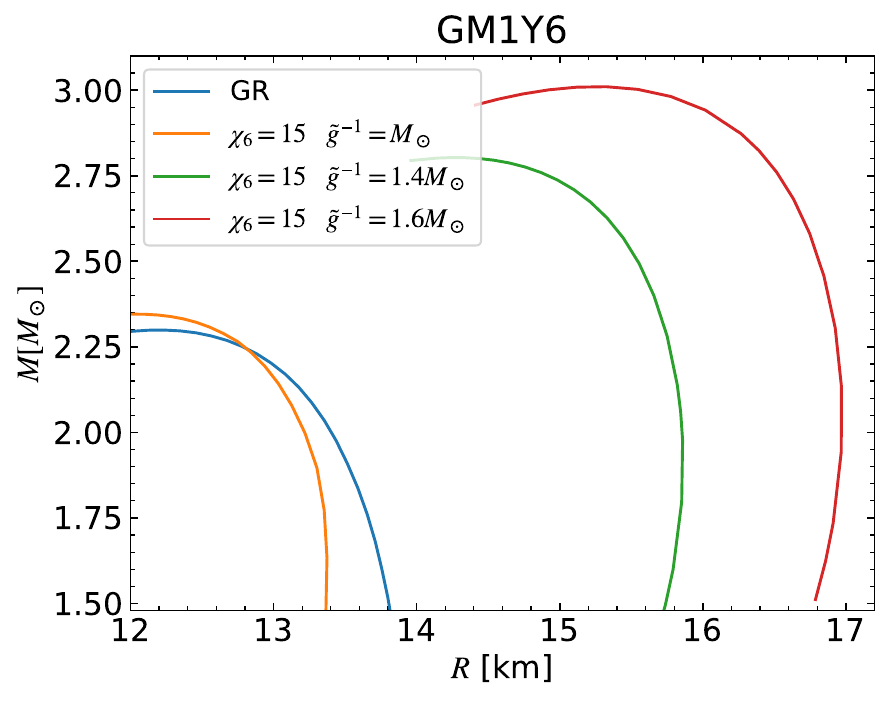}
    \caption{\textcolor{black}{gravitational mass versus radius of a NS whose baryonic matter is described by the EOS GM1Y6. We have fixed, in this case, the parameter $\chi_6=15$ for the minimal coupling scenario and varied the group coupling constant $\tilde{g}$ as indicated.}}
    \label{fig:massgap}
\end{figure}

\textcolor{black}{We have found out NS solutions with $\tilde{g}$ slightly less than $M_\odot^{-1}$ which are more massive than the previous NS solutions. The mass-radius equilibrium sequences with $\chi_6 = 15$, for the EOS GM1Y6 and two additional values of $\tilde{g}$, are shown in Fig. \ref{fig:massgap}.}
\subsection{Non-minimal coupling case}
\begin{figure*}
    \centering
    \includegraphics[width=0.49\textwidth]{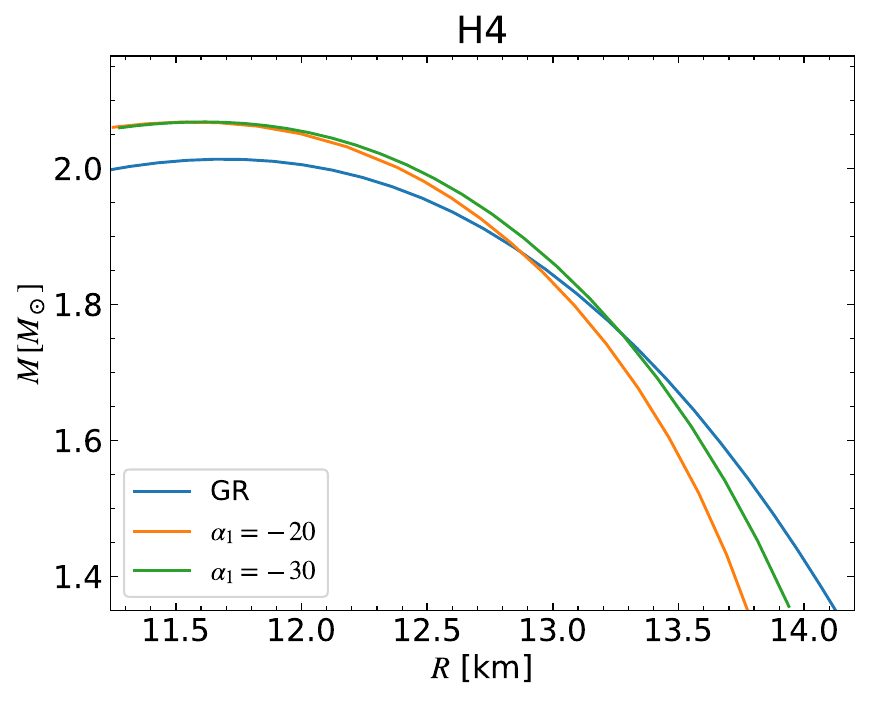}    
    \includegraphics[width=0.49\textwidth]{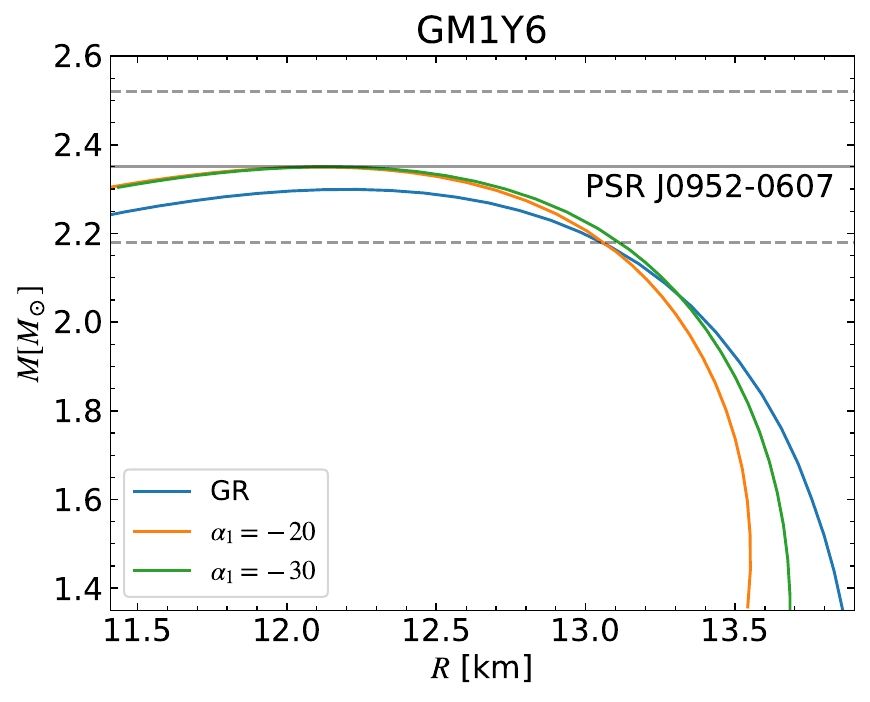}    
    \caption{Gravitational mass versus radius of a NS whose baryonic matter is described by the EOS H4 (left) and  GM1Y6 (right) for the non-minimal coupling case $\alpha_1$. The horizontal grey solid line in the right panel corresponds to the mass of the potentially heaviest observed NS PSR J0952-0607 ($2.35\pm 0.17 M_\odot$) \cite{Romani:2022jhd}, with the $1
\sigma$ uncertainty being represented by dashed grey lines.}
    \label{fig:MRA1}
\end{figure*}
We continue with the non-minimal coupling case, $\alpha_1\neq0$ and all other parameters vanishing. We have proceeded in a similar way to the minimal coupling case by first using a single polytropic EOS to determine the range for the parameter $\alpha_1$ so that the energy density of the vector fields does not dominate over the baryonic matter. We have found that to fulfill this requirement, $\alpha_1 \lesssim -10$.

\begin{figure*}
    \centering
    \includegraphics[width=0.95\linewidth]{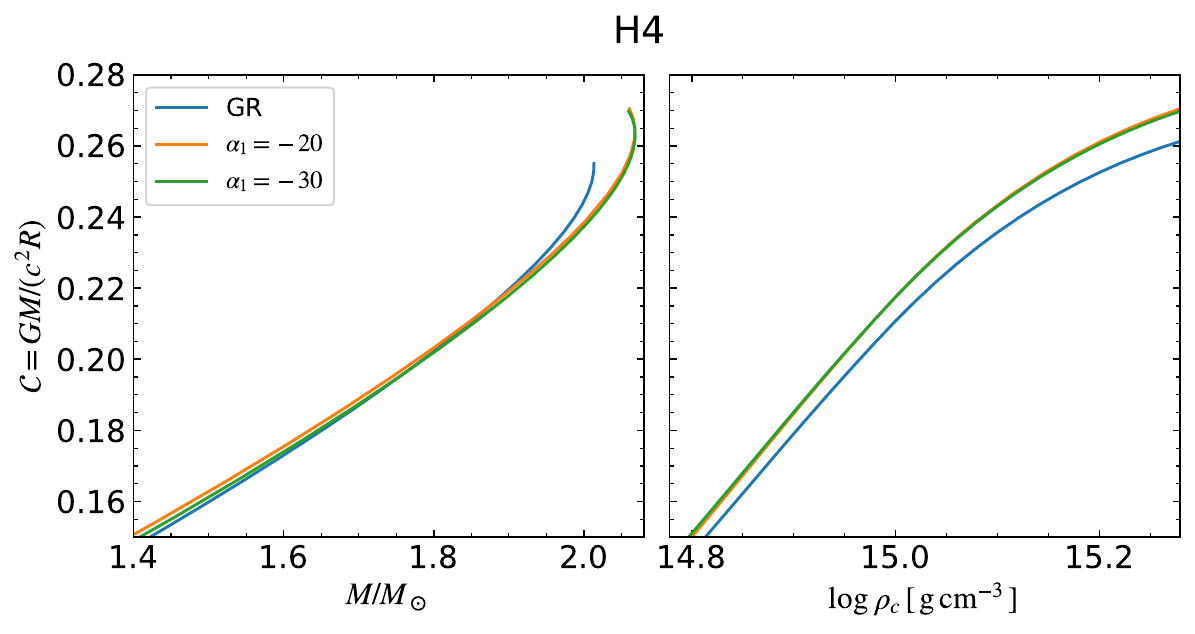}
    \caption{\textcolor{black}{Compactness of the H4 EOS equilibrium configurations shown in the left panel of Fig. \ref{fig:MRA1} as a function of the gravitational mass (left) and of the central baryonic mass density (right).}}
    \label{fig:CompactnessA1}
\end{figure*}

We have constructed equilibrium sequences of mass versus radius for the selected EOS which are shown in Fig. \ref{fig:MRA1}. The behaviour is similar to the minimal coupling case. The maximum mass is greater than in GR and the NSs are more compact than in GR as well; \textcolor{black}{this is clearly seen from Figs. \ref{fig:MRA1} and \ref{fig:CompactnessA1}.} However, the NSs are less compact than in the minimal coupling case. Again, the excess in gravitational mass comes from the contribution of the vector fields.

As mentioned before, the YM energy is dominant and creates very compact cores of baryonic matter. To construct a realistic NS model, there must be a counterterm in the full Lagrangian to cancel out the contribution of the YM term. The piece of the Lagrangian associated with $\alpha_1$ involves more terms and we have found in our solutions that its screening effect is greater than that of the simpler piece $\chi_6$.

\begin{figure}
    \centering
    \includegraphics[width=0.49\textwidth]{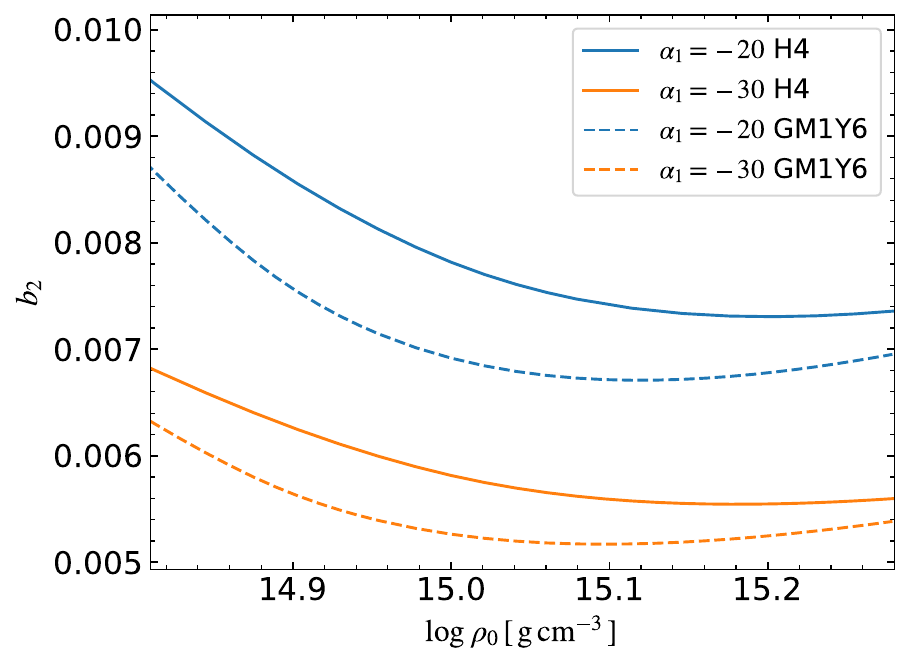}
    \caption{Shooting parameter (central energy density associated with the vector fields) versus central density of baryonic matter for different values of the coupling parameter and both EOSs as indicated in the legend.}
    \label{fig:b2-rho0-A1}
\end{figure}

The behaviour of the central energy density associated with the vector fields is shown in Fig. \ref{fig:b2-rho0-A1}. For both EOSs, when the baryonic matter decreases, the shooting parameter is seen to grow and the contribution to the gravitational mass from the vector fields becomes more significant. We have also found that, in general, the binding energy is positive, thus the configurations are potentially stable.

\section{Conclusions} \label{the conclusions}
\textcolor{black}{Ongoing efforts to describe compact objects in alternative theories of gravity have consolidated a promising programme to test gravity in the strong-field regime. Under this premise, we have examined in this work the effects of additional gravitational degrees of freedom on the baryonic matter of NSs within the GSU2P theory.} We have employed both analytical and numerical methods to solve the equations of structure for static and spherically symmetric NSs, with realistic EOSs, namely H4 and GM1Y6. \textcolor{black}{Although our results marginally depend on the EOS, which distinguishes high-density and low-density regions, they provide a preliminary but compelling approach to model realistic NSs in this alternative gravity theory.}

Within the framework of the EYM theory in the presence of baryonic matter, we initially observed that this type of solution exhibits a small core surrounded by a halo composed of vector fields. \textcolor{black}{This result raises the question of whether a similar object is formed in the GSU2P theory or whether a more extended baryonic dense core can be formed instead.} To elucidate this behaviour, we have analyzed two cases within the GSU2P theory: one featuring minimal coupling between the metric tensor and the vector fields, and the other with non-minimal coupling. For the purposes of this work, the additional vector gravitational degrees of freedom, i.e., those described by $B^{a}_{\mu}$, have been treated as a dark fluid.

\textcolor{black}{As a general trend, we have observed that the YM term tends to dominate the energy content when the free parameters are very small, approaching zero. This suggests a detailed exploration of the parameter space of the theory, for the cases where the dark fluid dominates the energy content. As a matter of fact,} we have explored the parameter space of $\chi_6$ and $\alpha_1$ to ensure that the YM energy density does not dominate over the contribution associated with baryonic matter. 
Specifically, we have determined the ranges $\chi_6\gtrsim 10$ and $\alpha_1\lesssim -10$, within which the YM contribution remains subdominant. \textcolor{black}{This allows for the formation of realistic NSs with a non-trivial vector field configuration surrounding the baryonic core.}

As a main conclusion of this work, we have found that NSs within the GSU2P theory exhibit greater compactness compared to those in GR with the same central baryonic mass density. This  effect is due to the energy contribution of the vector fields to the gravitational mass, which can be interpreted as the possible repulsive behaviour of the vector fields just as was shown in \cite{Martinez:2022wsy}. The first astrophysical implication of this result is the \textcolor{black}{plausible and} suitable explanation for very heavy NSs such as PSR J0952-0607. All these quantitative features can be described in both the minimal and non-minimal coupling cases. Moreover, these solutions enjoy the good property of stability, although based on the criteria of the binding energy which is a necessary but not a sufficient condition. This encourages employing more sophisticated methods based on perturbative arguments. This is a crucial aspect we are planning on addressing in a future work. 

It is worthwhile mentioning that GR is consistent to $1\sigma$ with the mass of PSR J0952-0607. \textcolor{black}{In fact, in GR, the theoretical upper limit on the NS maximum mass is $3.2~M_\odot$ \cite{1974PhRvL..32..324R}.  Accordingly, \citet{Annala:2021gom} studied a large ensemble of interpolated NS EOS incorporating  recent observational data from pulsar and gravitational wave events, concluding that a NS  could have a mass of up to $2.53~M_\odot$ in GR. Additionally, effects like rotation or pressure anisotropy could increase the NS maximum mass by around 20\%  (see e.g. \cite{2018ApJ...852L..25R,2024PhRvD.109d3025B}). These limits might explain the masses of compact objects near the lower boundary of the mass-gap, such as  the secondary  object of the merger event in GW190814 \cite{LIGOScientific:2020zkf},  which has a mass  between $2.5$ and $2.7~M_\odot$ (see e.g. \cite{2022ApJ...936...41L}). Future observations may place more stringent upper bounds on the NS maximum mass.}

\textcolor{black}{The next question we have addressed in this work is whether the theory can account for the mass gap. Interestingly, by properly fixing the group coupling constant $\tilde{g}$, the solution can be straightforwardly accommodated in the mass gap, providing a satisfactory explanation for the existence of massive compact objects beyond GR expectations. But, do these objects truly offer a purely gravitational explanation for the mass gap? Although this is a first step towards addressing this issue, it is necessary to explore the parameter space of the theory more comprehensively, along with a more rigorous treatment of stability. This will allow us to arrive at a more conclusive answer. On the other hand, whether there is a mass gap or not in the heart of GR is a subject that can shed light on the physics of supernova explosions, mass accretion of neutron stars, binary mergers, and potentially on the gravitational theory that governs these phenomena. Finally, the pressure of the vector fields in the t'Hooft-Polyakov configuration is anisotropic, so it may be interesting to analyze compact stars with an anisotropic EOS.}

\textcolor{black}{With the aid of current and future high-precision data, we aim at testing the consistency of this theory in other astrophysical scenarios and further constrain the untested parts of the theory. For example, gravitational wave observations from various astrophysical events can provide insights into any deviation from GR waveforms \cite{Yunes:2013dva,Perkins:2020tra,Moore:2021eok}. In addition, pulsar timing can reveal deviations in the timing residuals caused by the presence of fundamental (vector) fields \cite{NANOGrav:2023hvm,Zhang:2023lzt}. These examples, though speculative, offer stringent tests that may reveal the presence of fundamental massive vector fields around strong gravitational backgrounds in the form proposed in this paper.}

\section*{Acknowledgements} 

This research work has been funded by the Patrimonio Autónomo — Fondo Nacional de Financiamiento para la Ciencia, la Tecnología y la Innovación Francisco José de Caldas (MINCIENCIAS — COLOMBIA) under the grant No. 110685269447 RC-80740-465-2020, project 69553, by Universidad Industrial de Santander under the grant VIE 3921, and by Universidad Antonio Nari\~no under the grant VCTI 2024211. JFR is thankful for financial support from the Universidad Industrial de Santander, VIE, posdoctoral contract 004-4503 of 2024 with contractual registry No. 2024000737. 
We thank Jorge A. Rueda for discussions.

\appendix

\section{Equations of structure}\label{appES}
This appendix shows the explicit form of the equations of structure.
When there is \emph{only} a minimal coupling between the spacetime geometric structure and the vector fields, i.e. for $\alpha_1=\alpha_3=0$ in \eqref{eqn:lagrangian}, the equations are as follows:
\begin{widetext}
\begin{multline}
    m'+\frac{(2 m-r) w'^2 \left[\tilde{g}^2 r^2+(w+1)^2 \chi_{56}\right]}{\tilde{g}^4 r^3}+\frac{4 \chi _6 (w+1)^4}{\tilde{g}^4 r^4} \\+\frac{(w+1)^2 \left[(w+1)^2 \chi _{12}-\tilde{g}^2(w-1)^2-2 \mu^2\tilde{g}^2 r^2\right]}{2 \tilde{g}^4 r^2}= 4\pi r^2 \epsilon \,,\label{eqn:eqm-min}
\end{multline} 
\begin{equation}
    \delta '+\frac{2 w'^2 \left[\tilde{g}^2 r^2+(w+1)^2 \chi_{56}\right]}{\tilde{g}^4 r^3} + \frac{4\pi r^2 (P + \epsilon)}{r-2m}=0 \,,
\end{equation}
\begin{multline}
r (r-2 m) \left[(w+1)^2 \chi_{56}+\tilde{g}^2 r^2\right] w'' +r \chi_{56} (w+1) (r-2 m) w'^2-2 r \left[\tilde{g}^2
r^2+(w+1)^2 \chi_{56}\right]m' w'\\
-r(r-2 m) \left[\tilde{g}^2 r^2+(w+1)^2\chi_{56}\right]w' \delta ' +\left[2 m \tilde{g}^2 r^2+2 (w+1)^2 \chi_{56} (3 m-r)\right]w'\\
+r^2 (w+1) \left[\tilde{g}^2(w-\mu^2 r^2-w^2)+(w+1)^2 \chi _{12}\right]+8 \chi _6 (w+1)^3=0 \,,
\end{multline}
\end{widetext}
where $\chi_{12}\equiv 2\chi_1 +\chi_2$ and $\chi_{56} \equiv \chi _6- \chi _5$. As already mentioned in \cite{Martinez:2022wsy}, the parameter $\chi_4$ has no contribution to the field equations.

Now, for the \textcolor{black}{non-minimal coupling} case $\chi_1=\chi_2=\chi_4 =\chi_5=\chi_6 =0$ in  \eqref{eqn:lagrangian}, the respective equations are as follows:
\begin{widetext}
\begin{multline}
 \left\{1+\mathcal{C}_1\left[\frac{(w+1)}{\tilde{g}^2r}-2\frac{w'}{\tilde{g}^2}\right]\right\}m'- \frac{h}{\tilde{g}^2}\left(1-\mathcal{C}_2\right)w'^2-2 \left[h\,\mathcal{C}_3 - \mathcal{C}_1\frac{m}{r}\right]\frac{w'}{\tilde{g}^2}\\
+\frac{2r  h\,\mathcal{C}_1 }{\tilde{g}^2}w''
-\frac{\mu^2 (w+1)^2}{\tilde{g}^2}-\frac{\left(w^2-1\right)^2}{2 \tilde{g}^2r^2}+\frac{4 \left(25 \alpha _1-36 \alpha _3\right) (w+1)^4}{5 \tilde{g}^4r^4}\\
-\frac{\left(880 \alpha _1-1003 \alpha _3\right) m (w+1)^4}{15 \tilde{g}^5r^5}=4\pi r^2 \epsilon \, , 
\end{multline}
\begin{multline}
     h\,\left[1+\frac{\mathcal{C}_1(1+w)}{r}\right]\frac{\Phi'}{\tilde{g}^2} -  h\,\left(1-\mathcal{C}_4\right)\frac{w'^2}{\tilde{g}^2 r} - \frac{2 h\,\mathcal{C}_1}{\tilde{g}^2}\Phi'w'-2 h\mathcal{C}_5\frac{w'}{\tilde{g}^2}-\frac{m}{r^2}+\frac{\left(w^2-1\right)^2}{2 \tilde{g}^2 r^3}\\+\frac{\mu^2 (w+1)^2}{\tilde{g}^2 r}+\frac{\left(140 \alpha _1-101 \alpha _3\right) (w+1)^4}{5 \tilde{g}^4 r^5}-\frac{2 \left(100 \alpha _1-89 \alpha _3\right) m (w+1)^4}{5 \tilde{g}^4r^6}= 4\pi r P \,,
\end{multline}
\begin{multline}
    rh\,\mathcal{C}_1\left(\Phi''+\Phi'^2\right) - \mathcal{C}_1\Phi'm' +h\left(1-\mathcal{C}_4\right)\Phi'w'+ h\left(r\mathcal{C}_5+\mathcal{C}_1\frac{m}{r}\right)\Phi' - \mathcal{C}_5m'\\
    +\left(\frac{\mathcal{C}_4-1}{r}\right)m' w' + h\, (1-\mathcal{C}_4)w'' - h\,\mathcal{C}_4 w'^2 +\left[2h\,\mathcal{C}_4+\frac{m}{r}\left(1-\mathcal{C}_4\right)\right]\frac{w'}{r}\\
    -\mu^2 (w+1)-\frac{w \left(w^2-1\right)}{r^2} -\frac{\left(8 \alpha _1+3 \alpha _3\right) (w+1)^3}{\tilde{g}^2 r^4}-\frac{\left(16 \alpha _1-19 \alpha _3\right) m (w+1)^3}{3 \tilde{g}^2 r^5}=0\, ,\label{eqn:eqw-nonmin}
\end{multline}
\end{widetext}
where
\begin{align}
    \mathcal{C}_1 &\equiv\frac{\left(40 \alpha _1-67 \alpha _3\right)  (w+1)^3}{15 \tilde{g}^2r^3}\, ,\\
    \mathcal{C}_2&\equiv\frac{\left(85 \alpha _1-124 \alpha _3\right) (w+1)^2}{5\tilde{g} r^2}\, ,\\
    \mathcal{C}_3&\equiv\frac{\left(320 \alpha _1-407 \alpha _3\right) (w+1)^3}{15 \tilde{g}^2r^3}\, ,\\
    \mathcal{C}_4&\equiv\frac{\left(\alpha _1+2 \alpha _3\right) (w+1)^2}{\tilde{g}^2r^2}\, ,\\
    \mathcal{C}_5&\equiv\frac{ \left(160 \alpha _1-139 \alpha _3\right)  (w+1)^3}{15 \tilde{g}^2 r^4}\, .
\end{align}
\color{black}
\section{The Komar mass}\label{sec:appKomarADM}
The Komar mass in a stationary spacetime is analogous to the Newtonian concept of gravitational mass and is related to the local force required to hold a unit test mass in place in the presence of a gravitational field. The Komar mass is defined using the time-like Killing vector which, for the metric corresponding to the line element \eqref{eqn:metricII}, is given by
\begin{equation}
    \xi^{\mu}= (1,0,0,0).
\end{equation}
The Komar mass is given by the surface integral
\begin{equation}
    M \equiv -\frac{1}{8\pi} \lim_{S\to\infty}\int_S\nabla^{\mu}\xi^{\nu} dS_{\mu\nu},
\end{equation}
where $S$ is a 2-sphere.

For the case in study, the integrand reduces to
\begin{equation}
    \nabla^{\mu}\xi^{\nu}dS_{\mu\nu} = -2e^{2\Phi}(r-2m)^2 \Phi' r^2 \sin\theta \, d\theta\, d\phi.
\end{equation}
The metric functions behave as $\Phi = \Phi_{\infty} - m_{\infty}/r + \mathcal{O}(r^{-2}),\, m=m_{\infty} + \mathcal{O}(r^{-1})$ when $r\to \infty$ \cite{Martinez:2022wsy}. Therefore,
\begin{equation}
    M = \lim_{r\to\infty} e^{2\Phi}(r-2m)^2 \Phi' r^2 = e^{2\Phi_{\infty}}m_{\infty}.
\end{equation}
The value of $\Phi_{\infty}$ is chosen so that the time coordinate corresponds to the proper time measured by an observer at infinity, i.e. $e^{2\Phi_{\infty}}=1$. Given this normalization, the Komar mass can be expressed as stated in Eqs. \eqref{km} and \eqref{eqn:m-asymp-infty}. Note that the Komar mass depends only on the asymptotic values of the metric functions.
\section{Boundary terms of the gravitational action}\label{sec:GYH}
For a manifold with boundary, the gravitational variational problem is formulated by imposing that the gravitational fields vanish at the boundary, $\delta g_{\mu\nu}\rvert_{\partial \mathcal{M}} = \delta B_\mu^a\rvert_{\partial \mathcal{M}} = 0$.  Since the action contains second-order derivatives of the metric tensor, it is necessary to introduce a counterterm in order to eliminate the variations of the first-order derivatives of the metric tensor at the boundary, ensuring a well-posed problem. In GR, the Gibbons-York-Hawking boundary term gives a well-posed variational problem \cite{1977PhRvD..15.2752G,1996CQGra..13.1487H}. 
Besides, this term provides the correct Hamiltonian to reproduce the ADM mass.
This boundary term is given by
\begin{equation}
    S_{B}^{\rm GYH} \equiv \frac{1}{8\pi}\int_{\partial \mathcal{M}} \epsilon K \sqrt{h}d^3y, 
\end{equation}
where $\epsilon$ is +1 when $\partial \mathcal{M}$ is timelike and $-1$ when $\partial \mathcal{M}$ is spacelike, $K$ is the trace of the extrinsic curvature, $h$ is the determinant of the induced metric on the boundary, and $y^i$ are the coordinates on the boundary. 

In the GSU2P theory, boundary terms are necessary when second-order derivatives of the gravitational fields appear in the action,  as happens for the curvature couplings in the Lagrangian pieces \eqref{eqn:L43}-\eqref{eqn:L46}. For example, the boundary term corresponding to the non-minimal Lagrangian piece \eqref{eqn:L44} is
\begin{multline}
        S_{B}^{\rm GSU2P} = \frac{1}{8\pi}\int_{\partial \mathcal{M}} \Big[\left(B^{a}\cdot B_{a}\right)\left(B^{b}\cdot B_{b}\right)\\
        +2\left(B^{a}\cdot B^{b}\right)\left(B_{a}\cdot B_{b}\right)\Big]\epsilon K \sqrt{h}d^3y.
\end{multline}
Since $B_\mu^a$ generally does not vanish in asymptotically flat solutions, the Hamiltonian and ADM mass are modified. 
On the other hand, when the vector fields couple minimally to the metric tensor, no additional boundary terms to the Gibbons-York-Hawking one are needed because only $B_\mu^a$ and its first-order derivatives appear in the action\footnote{The theory was specifically designed to ensure this feature in order to avoid the Ostrogradsky instability (see \cite{GallegoCadavid:2022uzn,GallegoCadavid:2020dho}).}.A detailed Hamiltonian analysis will be addressed in a future work.

\color{black}
\section{Coefficients of the asymptotic series near the origin}\label{sec:appSeries}
This appendix presents the values of the non-vanishing coefficients of the asymptotic series near the origin \eqref{eqn:mseries0}-\eqref{eqn:Pseries}, obtained by solving the equations of structure shown in Appendix \ref{appES}:
\begin{equation}
a_3=\frac{4 \pi  \hat{\epsilon} _0}{3}+2b_2^2 \,,
\label{eqn:a3}
\end{equation}
\begin{multline}
    a_5=-\frac{8}{15} \pi ^2 \Xi  \left(\hat{P}_0+\hat{\epsilon} _0\right) \left(3 \hat{P}_0+\hat{\epsilon} _0\right)\\
    -\frac{8}{15}  \pi   \left[(3 \Xi -2) \hat{\epsilon }_0+3 (\Xi +2) \hat{P}_0\right]b_2^2+\frac{3 \mu^2}{5}b_2^2 \\
    -\frac{8 b_2^3}{5}+\left(\frac{4 \alpha_1}{3}+\frac{154 \alpha_3}{15}-4 \chi_6\right)b_2^4 \,,
\label{eqn:a5}
\end{multline}
\begin{multline}
   b_4 = \frac{1}{10}  \left(\mu ^2-8 \pi  \hat{P}_0+8 \pi  \hat{\epsilon }_0\right)b_2\\
   -\frac{3 b_2^2}{10} + \left[\alpha_1+\frac{7 \alpha_3}{10}+\frac{1}{5} (\chi_5-5 \chi_6+4)\right]b_2^3 \,,
\end{multline}
\begin{equation}
c_2=\frac{2}{3} \pi  \left(3 \hat{P}_0+\hat{\epsilon }_0\right)+2 b_2^2\,,
\label{}
\end{equation}
\begin{multline}
    c_4 = \left[6 \alpha _1-\frac{31 \alpha _3}{20}-\frac{2}{5} \left(\chi _5+5 \chi _6-6\right)\right]b_2^4\\
    -\frac{4 b_2^3}{5}+ \left\{\frac{\mu ^2}{5}-\frac{2}{5} \pi  \left[(\Xi -5) \hat{\epsilon }_0+(\Xi +3) \hat{P}_0\right]\right\}b_2^2 \\
    -\frac{2}{45}  \pi ^2 \left(3 \hat{P}_0+\hat{\epsilon }_0\right) \left[(3 \Xi -5) \hat{\epsilon }_0+3 (\Xi +5) \hat{P}_0\right]\,,
\end{multline}
\begin{equation}
    d_2 = -\frac{2}{3} \pi  \left(P_0+\epsilon _0\right) \left(3 P_0+\epsilon _0\right)-2 b_2^2 \left(P_0+\epsilon _0\right) \,,
\end{equation}
\begin{multline}
    d_4=\frac{4}{5} b_2^3 \left(\hat{P}_0+\hat{\epsilon }_0\right) -6  \left(\hat{P}_0+\hat{\epsilon }_0\right)\alpha _1 b_2^4\\
    +\frac{1}{20} b_2^4 \left(\hat{P}_0+\hat{\epsilon }_0\right) \left[31 \alpha _3+8 \left(5 \Xi +\chi _5+5 \chi _6-1\right)\right]\\
    +\frac{4}{45} \pi ^2 \left(\hat{P}_0+\hat{\epsilon }_0\right) \left(3 \hat{P}_0+\hat{\epsilon }_0\right) \left[4 \Xi  \hat{\epsilon }_0+3 (3 \Xi +5) \hat{P}_0\right]\\
    + \frac{2}{15} \pi  \left(\hat{P}_0+\hat{\epsilon }_0\right) \left[(13 \Xi -5) \hat{\epsilon }_0+(33 \Xi +39) \hat{P}_0\right]b_2^2\\
    -\frac{1}{5} \mu ^2 \left(\hat{P}_0+\hat{\epsilon }_0\right)b_2^2 \,,
\end{multline}
where 
$\Xi \equiv \tfrac{d\epsilon}{dP}(r=0)$.
\bibliography{References/referencias,References/references_post,References/references,References/references2,References/referencesPhD,References/sample,References/gcn, References/eos}
\end{document}